\documentclass[twocolumn,showpacs,aps]{revtex4}
\usepackage[dvips]{graphicx}
\usepackage{bm}

\begin{document}
\date{\today}
\title{Dissipation Under the Magnetic Field on 2D Surface}
\author{Naohisa Ogawa~$\dag$ and Shu-ichi Nagasawa~*}\footnotemark
\address{$\dag$ Department of Liberal Arts, Hokkaido Institute of Technology\\
006-8585 Sapporo, Japan\\
* Hakodate National College of Technology, Tokura-cho, 042-8501 Hakodate, Japan\\}

\begin{abstract}
It is well known that the convection of liquid solution or melt liquid 
is inhibited by the magnetic field experimentally,
 and so that we can make the larger and better crystals 
at the cost of slow growth rate \cite{Thompson}, \cite{Kasuga}, \cite{Kakimoto}.
We shall show here another two effects due to the magnetic field.
First, when the ionised molecule diffuses on the surface of crystal under the magnetic field 
applied normal to the surface,its diffusion constant decreases due to the magnetic effect.
Second, even if the molecule has no electric charge, if it has electric dipole moment, 
the kinetic energy decreases by different mechanism.
\end{abstract}
\maketitle

\stepcounter{footnote}\footnotetext{ogawanao@hit.ac.jp, ~~nagasawa@hakodate-ct.ac.jp}

\section{charged particle diffusion}
Let us consider the motion of charged particle on flat surface under the magnetic field.
Without magnetic field, there are various driving forces to make a motion of particle,
such as, random force due to temperature, fluid mechanical force, e.t.c..
We introduce the velocity due to these forces by $\vec{v}$.

\begin{equation}
\vec v = \mu \vec{F}_{ext},
\end{equation}

where $\mu$ is the mobility.
Then the electric current is given by

\begin{equation}
\vec{J}_{tot} = en\vec{v} + \vec{J}_{mag}.
\end{equation}

Where $e$ is the electric charge, $n$ is the particle density, 
and $\vec{J}_{mag}$ is the current induced by magnetic field, such as,

\begin{equation}
\vec{J}_{mag}= (\frac{\sigma}{en}) (\vec{J}_{tot} \times \vec{B}),
\end{equation}

where $\sigma$ is electric conductivity.
By mixing these two equations, we obtain
\begin{equation}
\vec{J}_{tot} = en\vec{v} + (\frac{\sigma}{en}) (\vec{J}_{tot} \times \vec{B}).
\end{equation}

From the iteration method, we obtain

\begin{eqnarray}
\vec{J}_{tot} &=& en\vec{v} 
+ (\frac{\sigma}{en}) (\vec{J}_{tot} \times \vec{B})\nonumber\\
&=& en\vec{v} + \sigma (\vec{v} \times \vec{B}) 
- (\frac{\sigma^2}{en})B^2 \vec{v} + {\cal O}(B^3).
\end{eqnarray}
The third term means the inhibition of motion due to the magnetic field.
\\

To solve the full equation without approximation, we set the coordinates as,

\begin{equation}
\vec{J}_{tot}=(J_x, J_y, 0), ~~~\vec{B}=(0, 0, B).
\end{equation}

Then we have

\begin{eqnarray}
\left[ 
\begin{array}{cc}
1&-\bar{\sigma}B  \\
\bar{\sigma}B &1  \\
\end{array} 
\right]
\left[ 
\begin{array}{c}
J_x  \\
J_y  \\
\end{array} 
\right]
= en 
\left[ 
\begin{array}{c}
v_x  \\
v_y  \\
\end{array} 
\right],
\end{eqnarray} 

where $\bar{\sigma}=\sigma/(en)$.
By multiplying inverse matrix in both hand sides, we obtain

\begin{equation}
J_x=\frac{en}{1+\bar{\sigma}^2 B^2}(v_x+\bar{\sigma}Bv_y),~~~
J_x=\frac{en}{1+\bar{\sigma}^2 B^2}(v_y-\bar{\sigma}Bv_x).
\end{equation}

If we use mobility instead of electric conductivity,
\begin{equation}
\sigma = e^2 n \mu,
\end{equation}
we obtain
\begin{equation}
\vec{J}_{tot}=\frac{en\vec{v}}{1+(e \mu B)^2}+ (e^2 n \mu) \frac{\vec{v} 
\times \vec{B}}{1+(e \mu B)^2}.
\end{equation}

Next we consider the case of diffusion.
We simply exchange the variable as
\begin{equation}
-D \nabla n= n \vec{v}.
\end{equation}

Then we have
\begin{eqnarray}
J_x &=& -\frac{eD}{1+(e \mu B)^2}(\partial_x n +e \mu B \partial_y n),\\
J_x &=& -\frac{eD}{1+(e \mu B)^2}(\partial_y n -e \mu B \partial_x n).
\end{eqnarray}

By using the charge conservation law

\begin{equation}
- e \frac{\partial n}{\partial t} = div (\vec{J}_{tot}),
\end{equation}

we obtain the diffusion equation.

\begin{equation}
\frac{D}{1+(e \mu B)^2}\nabla^2 n =  \frac{\partial n}{\partial t}.
\end{equation}

This shows the new effective diffusion constant.
\begin{equation}
D \to D' \equiv \frac{D}{1+(e \mu B)^2}.
\end{equation}

In this model, energy dissipation is induced by 
the existence of electric conductivity or mobility.
The energy loss is easily calculated as,
\begin{equation}
\Delta E = \frac{\rho}{2}\{(\vec{J}_{tot}/(en))^2-(\vec{v})^2\}
=-\frac{(e \mu B)^2}{1+(e \mu B)^2}E.
\end{equation}

\section{motion of a particle with electric dipole moment under magnetic field}

Next we consider the motion of electric dipole moment which is planar to the surface 
under the magnetic field applied normal to the surface.
The distance of two charges are taken to be constant length $l$.
Then the relative and center of mass coordinates are
\begin{equation}
\vec{r}=\vec{x}_1 - \vec{x}_2=l \vec{n}, ~~~ \vec{R}=\frac{\vec{x}_1 + \vec{x}_2}{2}.
\end{equation}
respectively.

\begin{figure}
\centerline{\includegraphics[width=3cm]{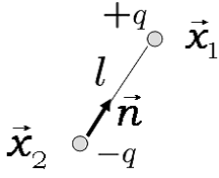}}
\caption{Electric dipole moment}
\end{figure}

The motion of equations are
\begin{equation}
m \ddot{\vec{x}}_1 =  e \dot{\vec{x}}_1 \times \vec{B} + \vec{n}f,~~~
m \ddot{\vec{x}}_2 = -e \dot{\vec{x}}_2 \times \vec{B} - \vec{n}f,
\end{equation}
where $f$ is the constraint force.

From these two equations, we obtain
\begin{eqnarray}
\ddot{\vec{r}} &=&  \frac{2e}{m} \dot{\vec{R}} \times \vec{B} + \frac{2f}{m} \vec{n}, \label{eq:relative}\\
\ddot{\vec{R}} &=&  \frac{e}{2m}  \dot{\vec{r}} \times \vec{B}.\label{eq:mass_center}
\end{eqnarray}

First we consider the relative motion.\\

We obtain from (\ref{eq:relative})

\begin{equation}
\ddot{\vec{n}} =  \frac{2e}{ml} \dot{\vec{R}} \times \vec{B} + \frac{2f}{ml} \vec{n}.\label{eq:nom}
\end{equation}

By multiplying $\dot{\vec{n}}$ in both hand sides of (\ref{eq:nom}), we obtain

\begin{equation}
\ddot{\theta} =  -\frac{2eB}{ml} (\dot{Y} \sin \theta + \dot{X} \cos \theta),\label{eq:rot1}
\end{equation}

where
\begin{eqnarray}
\vec{n}&=&(\cos \theta, \sin \theta),~~~
\dot{\vec{n}}=(-\sin \theta, \cos \theta) \dot{\theta},\nonumber\\
\ddot{\vec{n}}&=&(-\sin \theta, \cos \theta) \ddot{\theta} 
- (\cos \theta, \sin \theta) \dot{\theta}^2,\nonumber
\end{eqnarray}

and $\vec{R}=(X,Y)$ are utilised.\\

On the other hand, by multiplying $\vec{n}$ in both hand sides of (\ref{eq:nom}), we obtain
\begin{equation}
f =  -\{ \frac{ml}{2}\dot{\theta}^2 + eB (\dot{Y} \cos \theta - \dot{X} \sin \theta)\}.
\end{equation}
This equation simply determines constraint force $f$.

Next we consider the motion of center of mass.
From (\ref{eq:mass_center}), we obtain
\begin{equation}
\ddot{X}=\frac{eBl}{2m} \dot{\theta} \cos \theta, ~~~~\ddot{Y}=\frac{eBl}{2m} \dot{\theta} \sin \theta.
\end{equation}

The time integration gives
\begin{equation}
\dot{X}=\frac{eBl}{2m} \sin \theta + U_x, ~~
~~\dot{Y}=-\frac{eBl}{2m} \cos \theta + U_y,\label{eq:motion}
\end{equation}

where $U_x,~~U_y$ are the integration constants.

By putting (\ref{eq:motion}) into (\ref{eq:rot1}), we obtain
\begin{equation}
\ddot{\theta} =  -\frac{2eB}{ml} (U_y \sin \theta + U_x \cos \theta).
\end{equation}

Now let us denote

\begin{equation}
U_x = U \cos \phi, ~~~ U_y = U \sin \phi.
\end{equation}

Then we have equation

\begin{equation}
\ddot{\theta} =  -\frac{2eB}{ml}U \cos (\theta - \phi).\label{eq:rot}
\end{equation}

This is nothing but the equation of pendulum.\\
To see it clearly, we write down the effective energy of this local system by using $p=el$.

\begin{equation}
E_{eff} = \frac{ml^2}{4}\dot{\theta}^2  + pB U \{\sin (\theta - \phi)+1\}. \label{eq:eff}
\end{equation}

\begin{figure}
\centerline{\includegraphics[width=6cm]{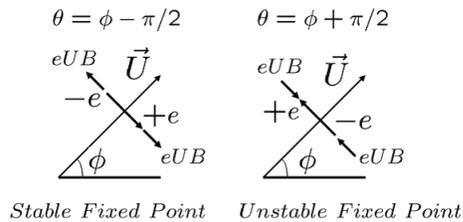}}
\caption{Fixed points}
\end{figure}

Compare to the pendulum weight, the lowest point corresponds to the stable fixed point,
 which is shown in figure 2 (left). 
On the other hand, the highest point corresponds to the unstable fixed point.
Note that when $B<0$ this relation is exchanged.

\section{Energy conservation and Mechanism of dissipation}

Let us  calculate the kinetic energy of center of mass.

\begin{eqnarray}
E_{kin} &=& \frac{M}{2}(\dot{X}^2+\dot{Y}^2) \nonumber\\
&=& \frac{M}{2}((\frac{pB}{M} \sin \theta + U_x)^2+(-\frac{pB}{M} \cos \theta + U_y)^2)\nonumber\\
&=&\frac{(pB)^2}{2M}+\frac{M U^2}{2} + pB (\sin \theta ~U_x -\cos \theta ~U_y)\nonumber\\
&=& const. + pBU \sin (\theta - \phi), \label{eq:kin}
\end{eqnarray}
where $M=2m$, the mass of dipole moment.
Therefore
\begin{equation}
\frac{d E_{kin}}{dt} = pBU \dot{\theta} \cos (\theta - \phi).
\end{equation}

By using (\ref{eq:rot}), we obtain
\begin{equation}
\frac{d E_{kin}}{dt} = -\frac{p^2M}{4e^2} \frac{d}{dt}(\frac{\dot{\theta}^2}{2}).
\end{equation}

We define the moment of inertia $I$ as
\begin{equation}
I \equiv \frac{Ml^2}{4}.
\end{equation}

Then we have
\begin{equation}
\frac{d}{dt}(E_{kin} + \frac{I}{2} \dot{\theta}^2) = 0.
\end{equation}

The total energy is the sum of kinetic energy and rotating energy that conserves.
The key point here is, when the magnetic field is absent, each energy conserves independently.
But when the field applied, relative and center of mass motion couples.
Therefore when the relative motion is put in heat bath and give its energy to the bath, 
total energy decrease and we find the dissipation not only the relative motion
but also the motion of center of mass.
This mechanism gives the effect of dissipation 
to the motion of center of mass only when magnetic field applied.
To realise the idea, we give a model in the next section.

\section{Model of dissipation}

We introduce the dissipation only to the relative motion, 
and see how the kinetic energy of center of mass decreases when the magnetic field applied.

Then our equation for the relative (rotating) motion is described by
\begin{equation}
\ddot{\vec{r}} =  \frac{2e}{m} \dot{\vec{R}} \times \vec{B} + \frac{2f}{m} \vec{n}-\frac{\alpha}{l}\dot{\vec{r}} \label{eq:relative2}
\end{equation}
instead of (\ref{eq:relative}) with positive $\alpha$.

And that gives 
\begin{equation}
\ddot{\theta} =  -\frac{2eB}{ml}U \cos (\theta - \phi) -\alpha \dot{\theta}, \label{eq:rot3}
\end{equation}
instead of (\ref{eq:rot}).

Then we obtain the interesting relation.
\begin{equation}
\frac{d E_{eff}}{dt}= \frac{d}{dt} ( E_{kin}+ \frac{I}{2}\dot{\theta}^2)=-\alpha I \dot{\theta}^2 \leq 0.
\end{equation}

That is, $E_{eff}$ is the dissipating total energy.
From this dissipating property and from the form of $E_{eff}$, 
we see $\theta \to \phi \mp \pi/2$ when $t \to \infty$.
The signature corresponds to the one of magnetic field, i.e. $B = \pm \mid B \mid$.
Then the kinetic energy $E_{kin}$ takes minimum value at  $t \to \infty$.
\begin{eqnarray}
E_{kin} \to E_{0} &=& \frac{(pB)^2}{2M}+\frac{M U^2}{2} -  p \mid B \mid U \nonumber\\
&=& \frac{M}{2}(U-  \frac{p \mid B \mid}{M})^2. \label{eq:energy}
\end{eqnarray}
Therefore the kinetic energy decreases due to the dissipation of relative (rotational) motion 
only when the magnetic field applied.
Furthermore, it should be noted that the minimum kinetic energy is Non zero unless
\begin{equation}
U= \omega_{c} l,
\end{equation}
holds.  Where the Lamor frequency $\omega_{c}$ is defined by
\begin{equation}
\omega_{c} \equiv \frac{e \mid B \mid}{M}.
\end{equation}
From the existence of specific length $l$ we have a constant $\omega_{c}l$ with  dimension of velocity .
Then some kind of ``resonance'' occurs with velocity $U$ 
that reduces the minimum kinetic energy to be zero.
\\

The simulation of the motion of electric dipole is given in 3 figures. (figure 3-5)
Parameters are taken as $U=3$, $\phi = \pi$, $\theta(t=0)=\dot{\theta}(t=0)=0$, $\omega_{c} l=\pi$, and $\alpha =0.4$.
Figure 3 shows the time dependence of  (effective) total energy.
Figure 4 shows the time dependence of the kinetic energy with its mean value for each one period, and
Figure 5 shows the trajectory of the center of mass.

\begin{figure}
\centerline{\includegraphics[width=5cm]{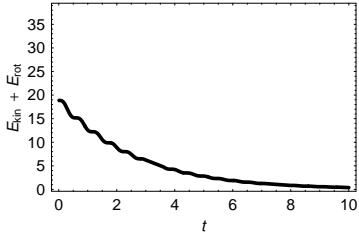}}
\caption{Total energy}
\end{figure}

\begin{figure}
\centerline{\includegraphics[width=5cm]{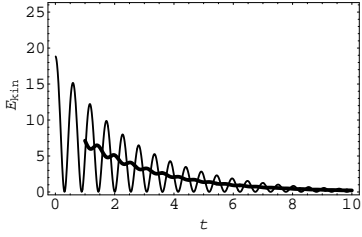}}
\caption{Kinetic energy of center of mass}
\end{figure}

\begin{figure}
\centerline{\includegraphics[width=5cm]{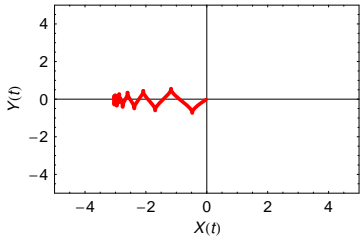}}
\caption{Trajectory of center of mass}
\end{figure}

From these result, our scenario that the kinetic (mean) energy dissipates when the magnetic field applied, is realized.

\section{Thermal effect to our dipole system}

Next we put the system of relative motion in the heat bath.
Then since the energy transfers from the rotating motion to the heat bath, 
we may expect the similar dissipative dynamics occurs in the diffusion system.

By taking the ensemble average of  equation (\ref{eq:motion}), we have
\begin{eqnarray}
<v_x> &=& +\frac{p\mid B \mid}{M} <\sin \theta>  +  U_x ,\\
<v_y> &=& -\frac{p\mid B \mid}{M} <\cos \theta>  +  U_y.
\end{eqnarray}

This average is re-interpreted as the thermal average due to the Ergode hypothesis as follows.

\begin{equation}
< f (\theta ) >=\frac{\int {d\theta } \int {dP_\theta } f(\theta )\exp [-\beta 
H(\theta ,P_\theta )]}{\int {d\theta } \int {dP_\theta } \exp [-\beta 
H(\theta ,P_\theta )]},
\end{equation}

with Hamiltonian

\begin{equation}
H=\frac{1}{ml^2}P_\theta ^2 + pB U \{ \sin (\theta -\phi)+1 \}.
\end{equation}

In the zero temperature limit $\beta \to \infty$, we have the lowest potential energy, that is,

\begin{equation}
 <\sin \theta > \simeq -\cos \phi, ~~~<\cos \theta > \simeq  \sin \phi,
 \end{equation}
 for $B>0$. And 

\begin{equation}
 <\sin \theta > \simeq \cos \phi, ~~~<\cos \theta > \simeq -\sin \phi,
\end{equation}
for $B<0$.

Therefore 
\begin{eqnarray}
 <v_x> &=& - \frac{p \mid B \mid}{M}\cos \phi + U_x  = \{ 1 - \frac{p \mid B \mid}{MU} \} U_x ,~~ \label{eq:0tempx}  \\
 <v_y> &=& - \frac{p \mid B \mid}{M}\sin \phi  + U_y  = \{ 1- \frac{p \mid B \mid}{M U} \} U_y .~~ \label{eq:0tempy}
 \end{eqnarray}
 
The mean velocity becomes zero when 

$$\frac{p \mid B \mid}{M U} = \frac{\omega_{c}l }{U }= 1,$$

this is related to the case $E_{kin}=0$ as shown in equation (\ref{eq:energy}).
This is NOT a dissipation, but is some ``resonance'' as we discussed already.

The finite temperature case can be obtained by explicit calculation as well.

\begin{eqnarray}
<\sin \theta > &=& \frac{\int {d\varepsilon } \sin (-\frac{\pi 
}{2}+\varepsilon +\phi )\exp [-\alpha (1-\cos \varepsilon )]}{\int {d\varepsilon } \exp [-\alpha (1-\cos \varepsilon)]}\nonumber\\
&=& -\cos \phi \frac{\int {d\varepsilon } \cos \varepsilon 
\exp [\alpha \cos \varepsilon]}{\int {d \varepsilon } \exp [\alpha \cos \varepsilon]} \nonumber\\
&=&-\cos \phi \frac{d}{d\alpha }\log \{\int {d\varepsilon } 
\exp [\alpha \cos \varepsilon ]\}.
\end{eqnarray}
with 
$$\alpha =\beta BpU.$$

In the same way,

\begin{eqnarray}
<\cos \theta > &=& \frac{\int {d\varepsilon } \cos (-\frac{\pi 
}{2}+\varepsilon +\phi )\exp [-\alpha (1-\cos \varepsilon )]}{\int {d\varepsilon } \exp [-\alpha (1-\cos \varepsilon)]}\nonumber\\
&=& \sin \phi \frac{\int {d\varepsilon } \cos \varepsilon 
\exp [\alpha \cos \varepsilon]}{\int {d \varepsilon } \exp [\alpha \cos \varepsilon]} \nonumber\\
&=& \sin \phi \frac{d}{d\alpha }\log \{\int {d\varepsilon } 
\exp [\alpha \cos \varepsilon ]\}.
\end{eqnarray}

By using the integration form of 0th modified Bessel function

\begin{equation}
I_0 (\alpha) = \frac{1}{2\pi} \int \limits_{-\pi }^{+\pi } {d\varepsilon } \exp [\alpha \cos \varepsilon],
\end{equation}

we obtain the mean velocity
\begin{equation}
 <\vec{v}> = \{1- \frac{pB}{MU} \frac{I_0 ^\prime (\alpha)}{I_0 (\alpha)}\} \vec{U}  ,
\end{equation}

while
$ I_{0}'/I_{0}$ is the odd function of $\alpha$ and so the signature of $B$ does not change the form of function.

Note that from the asymptotic expansion of modified Bessel function, we obtain

\begin{equation}
\lim_{ \alpha \to \pm \infty}\frac{I_0'(\alpha)}{I_0(\alpha)} \pm 1.
\end{equation}

Therefore we have
\begin{equation}
 <\vec{v}>  \to   \{1- \frac{p \mid B \mid}{M U} \} \vec{U},
\end{equation}

in zero temperature limit, which is consistent with previous result (\ref{eq:0tempx}) , (\ref{eq:0tempy}) .

The thermal expectation value of kinetic energy of center of mass is given as

\begin{eqnarray}
<E_{kin}> &=& \frac{M}{2} <v_{x}^{2}+v_{y}^{2}> \nonumber\\
&=& E_{0} + M \omega_{c } l U \{1-\frac{I_{0}'(\alpha)}{I_{0}(\alpha)}\}.
\end{eqnarray}

Where
\begin{equation}
 E_{0} \equiv  \frac{M}{2} (U-\omega_{c}l)^{2}.
\end{equation}

The figure 6 shows $<E_{kin}> -E_{0}$ in the unit of $M \omega_{c } l U$   as the function of dimensionless temperature $T/T^{*}$, where
$$T^{*} \equiv \frac{\mid B \mid pU}{2k}.$$
This function is monotone increasing function  from 0 to 1 for  the temperature increases from 0 to $\infty$. Therefore the mean kinetic energy has the asymptotic values as

\begin{eqnarray}
<E_{kin}> &\to&  E_{0} = \frac{M}{2} (U-\omega_{c}l)^{2}~~for~T \to 0, \\
&\to& \frac{M}{2} U^{2} + \frac{M}{2} (\omega_{c}l)^{2}~~for~T \to \infty.~~
\end{eqnarray}

The reason why the kinetic energy is finite under the infinite temperature is as follows.
The kinetic energy is essentially the same as the periodic potential energy in $E_{eff}$ as shown in (\ref{eq:eff}) and (\ref{eq:kin}).
Therefore if the temperature increases and the energy of relative motion (pendulum) comes up, 
mean value of potential energy is always  finite and so as the kinetic energy.

This temperature dependence of kinetic energy is absent when the magnetic field disappeared.
Such a dependence shows the energy transition between center of mass motion, relative motion and heat bath as we have expected.
If the temperature of heat bath decreases,  the energy of relative motion transfers to the bath, and energy of center of mass transfers as well.

\begin{figure}
\centerline{\includegraphics[width=5cm]{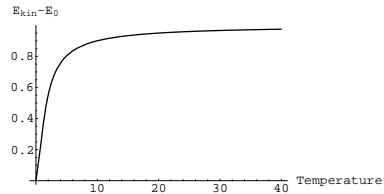}}
\caption{$<E_{kin}> -E_{0}$ in the unit of $M \omega_{c } l U$ 
as a function of dimensionless temperature $T/T^{*}$}
\end{figure}

\section{Summary}
Here we have shown the new mechanism that  the magnetic field connects two dynamical systems,
and if one system has the dissipation, another system becomes also dissipative only when the magnetic field is applied.
This mechanism might also work in many other physical systems under the magnetic field.
Another interesting point is that when the  electric dipole is placed in magnetic field, we have a constant with dimension of velocity.
Then some kind of ``resonance''  phenomenon occurs and we have critical value of magnetic field which makes lowest kinetic energy vanish.
\begin{equation}
B^{*} = \frac{MU}{el},
\end{equation}
where $U$ is a value of center of mass velocity when  the magnetic field is absent.

\end{document}